\documentclass[aps,prb,reprint,superscriptaddress,showpacs,showkeys]{revtex4-1}

\usepackage{graphicx}

\begin{document}


\title{$\mu$SR investigation of magnetically ordered states \\ in A-site ordered perovskite manganites $R$BaMn$_2$O$_6$ ($R$ = Y and La) }



\author{Y.\ Kawasaki}
\email{yu@pm.tokushima-u.ac.jp}
\affiliation{Institute of Technology and Science, The University of Tokushima, Tokushima 770-8506, Japan}
\author{T.~Minami}
\altaffiliation[Present address: ]{Seiryo Engineering Co., Ltd.}
\affiliation{Institute of Technology and Science, The University of Tokushima, Tokushima 770-8506, Japan}
\author{M.~Izumi}
\altaffiliation[Present address: ]{Nichia Corporation}
\affiliation{Institute of Technology and Science, The University of Tokushima, Tokushima 770-8506, Japan}
\author{Y.~Kishimoto}
\affiliation{Institute of Technology and Science, The University of Tokushima, Tokushima 770-8506, Japan}
\author{T.\ Ohno}
\affiliation{Institute of Technology and Science, The University of Tokushima, Tokushima 770-8506, Japan}
\author{K.H.\ Satoh}
\altaffiliation[Present address: ]{Quantum Design, Inc.}
\affiliation{Graduate University for Advanced Studies (SOKENDAI), Tsukuba 305-0801, Japan}
\author{A.~Koda}
\affiliation{Graduate University for Advanced Studies (SOKENDAI), Tsukuba 305-0801, Japan}
\affiliation{Muon Science Laboratory, Institute of Materials and Structure Science, High Energy Accelerator Research Organization (KEK), Ibaraki 319-1106, Japan}
\author{R.\ Kadono}
\affiliation{Graduate University for Advanced Studies (SOKENDAI), Tsukuba 305-0801, Japan}
\affiliation{Muon Science Laboratory, Institute of Materials and Structure Science, High Energy Accelerator Research Organization (KEK), Ibaraki 319-1106, Japan}
\author{J.L.\ Gavilano}
\affiliation{Laboratory for Neutron Scattering, PSI and ETHZ, Villigen CH-5232, Switzerland}
\author{H.\ Luetkens}
\affiliation{Laboratory for Muon-Spin Spectroscopy, Paul Scherrer Institut, Villigen CH-5232, Switzerland}
\author{T.\ Nakajima}
\affiliation{AMRI, National Institute of Advanced Industrial Science and Technology, Tsukuba 305-8565, Japan}
\author{Y.\ Ueda}
\affiliation{Institute for Solid State Physics, University of Tokyo, Kashiwa 277-8581, Japan}


\date{\today}

\begin{abstract}
The magnetically ordered states of the A-site ordered perovskite manganites LaBaMn$_2$O$_6$ and YBaMn$_2$O$_6$ have been investigated by muon spin relaxation in zero external magnetic field.
Our data reveal striking differences in the nature of the magnetically ordered state between these materials. 
For LaBaMn$_2$O$_6$, the $\mu$SR time-spectra in the ferromagnetic state below $\simeq 330$ K reveal a strongly inhomogeneous phase, reminiscent of a Griffiths phase.
Within this magnetically inhomogeneous phase, an antiferromagnetic state develops below 150 K, which displays well defined static internal magnetic fields, but reaches only 30\% of the volume fraction at low temperatures. 
A broad distribution of $\mu$SR relaxation rates is inferred down to the lowest temperatures. 
This behavior is similar to that in the A-site disordered La$_{0.5}$Ba$_{0.5}$MnO$_3$\@.
On the other hand, for YBaMn$_2$O$_6$, the $\mu$SR time spectra for both (i)  the charge and orbital ordered and (ii) the paramagnetic phases reveal rather homogeneous states, namely, an exponential relaxation in the paramagnetic state and well defined muon spin oscillation in the antiferromagnetic state.  

\end{abstract}


\maketitle


\section{Introduction}

In the last few decades, a great deal of experimental and theoretical research efforts have been devoted to perovskite manganites with chemical formulas of $R_{1-x}^{3+}A_x^{2+}$MnO$_3$ ($R^{3+}$ = rare earth ions, $A^{2+}$ = alkaline-earth ions) because they show a rich variety of fascinating electromagnetic properties, such as the colossal magnetoresistance (CMR), charge and orbital ordering (COO) and metal-insulator transition.
Recently, a great deal of attention is given to A-site-ordered perovskite manganites, materials with an ordered arrangement of two cations, $R^{3+}$ and $A^{2+}$, at the A site of the perovskite structure.\cite{alonso03, ueda04}

$R$BaMn$_2$O$_6$, an A-site-ordered manganite is formed by substituting 50\% of rare earth by barium ions in $R$MnO$_3$.\cite{millange98,troyanchuk02}
The crystal structure of $R$BaMn$_2$O$_6$ is schematically shown in Fig.~1\@.
It has a layered ordering of $R^{3+}$ and Ba$^{2+}$ ions along the $c$-axis, resulting in a stacking of two dimensional structures with the sequence of -$R$O-MnO$_2$-BaO-MnO$_2$- planes.
The structure of $R$BaMn$_2$O$_6$ at room temperature has a tetragonal $a_p\times b_p\times2c_p$ cell ($a_p=b_p$) with no tilt of MnO$_6$ octahedra for $R$ = La, Pr and Nd (dashed lines in Fig.~1), while for $R$ = Sm-Y the structure has a larger unit cell of $\sqrt{2}a_p\times\sqrt{2}b_p\times2c_p$ with a tilt of MnO$_6$ octahedra.\cite{nakajima02a,nakajima04jssc,millange98}
Here, the three lattice constants: $a_p$, $b_p$ and $c_p$ are associated with the cell of the corresponding A-site disordered $R_{0.5}$Ba$_{0.5}$MnO$_3$\@.\cite{ueda04}
The $\sqrt{2}a_p\times\sqrt{2}b_p\times2c_p$ unit cells for $R$ = Sm, Eu and Gd are orthorhombic but those for $R$ = Tb, Dy, Ho and Y are monoclinic.\cite{nakajima02,nakajima02a,nakajima04jssc,trukhanov02,uchida02}

The ordering of $R^{3+}$ and Ba$^{2+}$ ions dramatically changes the phase diagram of the disordered $R_{0.5}$Ba$_{0.5}$MnO$_3$\@.\cite{ueda04}
The remarkable features of the A-site ordered $R$BaMn$_2$O$_6$ are the following.
(i) The COO transition temperature $T_{\rm CO}$ is very high for $R$ = Sm-Y, while magnetic glassy states are common in $R_{0.5}$Ba$_{0.5}$MnO$_3$ except for $R=$ La\@.
(ii) In the phase diagram of temperature $T$ vs.\ the ratio of ionic radius $r_I$ ($= r_{R^{3+}}/r_{\rm Ba^{2+}}$), there is an abrupt phase transition at near $r_I =0.785$ between a COO and a ferromagnetic metal (FM) states with critical temperatures $T_{\rm CO}(r_I)$ and $T_C(r_I)$, respectively. 
Near the phase boundary no drastic changes in $T_C$ or $T_{\rm CO}$ are observed.
The second feature of $R$BaMn$_2$O$_6$ is considered to be a key characteristic for the development of new materials that exhibit CMR at room temperature.\cite{nakajima05,bhargava08}


Among the series of $R$BaMn$_2$O$_6$, YBaMn$_2$O$_6$ has the highest ordering temperatures; a structural change from a triclinic to a monoclinic at $T_t \simeq $ 520 K, charge and orbital order at $T_{\rm CO} \simeq$ 480 K and an antiferromagnetic spin order with $T_N$ just below 200 K\@.\cite{ueda04, akahoshi03, nakajima02, nakajima02a, kageyama03,nakajima04jssc,zakharov08,miyauchi10}
The distinct hysteresis at  $T_t$ and $T_N$ both in the susceptibility and the electrical resistivity indicate these transitions to be first order, while the transition at $T_{\rm CO}$ was suggested to be second order.\cite{nakajima04jssc}
These transition temperatures are by far higher than the spin-glass transition temperature of about 50 K in the disordered form, Y$_{0.5}$Ba$_{0.5}$MnO$_3$\@.
The high $T_{\rm CO}$ is considered to be due basically to the stacking order of Y$^{3+}$ and Ba$^{2+}$ in the structure, which results in the absence of electrostatic potential disorder.

\begin{figure}[tb]
\centering
\includegraphics[width=6.5cm]{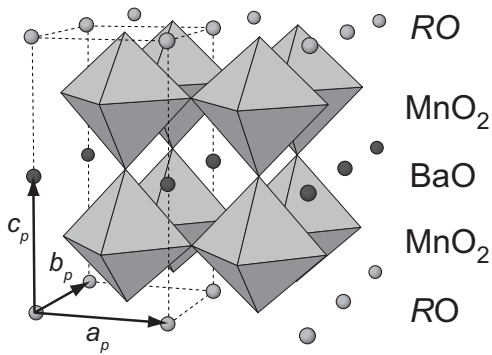}
\caption{Crystal structure of $R$BaMn$_2$O$_6$, where the three lattice constants: $a_p$, $b_p$ and $c_p$ are associated with the cell of the corresponding A-site disordered $R_{0.5}$Ba$_{0.5}$MnO$_3$\@.
Dashed lines indicate the tetragonal unit cell of $R$BaMn$_2$O$_6$ ($R$ = La, Pr and Nd). Distortions and tilting of MnO$_6$ octahedra are not shown.}
\end{figure}

The ordering patterns of orbital, charge and spin in YBaMn$_2$O$_6$ are also closely related to the peculiar structure to the A-site ordered perovskite.\cite{kageyama03}
The charge order below $T_{\rm CO}$ is suggested to be a checker-board type in the MnO$_2$ plane.
The orbital ordering pattern in the MnO$_2$ plane is the so-called charge-exchange (CE) type, which is the prototypical case in $R_{0.5}A_{0.5}$MnO$_3$ (for example, see Fig.\ 3A of reference\cite{tokura00}), but here it has a new stacking pattern with a fourfold periodicity along the $c$-axis, most probably $\alpha\alpha\beta\beta$ type, the same as in SmBaMn$_2$O$_6$\@.\cite{arima02,akahoshi04,fernandez08,morikawa12}
The rearrangement of the stacking pattern in the orbital order is induced by the spin order below $T_N$\@.
The spin ordering pattern in the MnO$_2$ plane is of the zigzag CE-type with the stacking pattern of $\alpha\alpha\beta\beta$ type along the $c$-axis.
The Y/Ba order along the $c$-axis in YBaMn$_2$O$_6$ may induce alternating ferro- and antiferromagnetic interactions between neighboring layers, resulting in the observed spin structure.
Hereafter, we denote this antiferromagnetic insulating phase as AFI(CE) phase.

A related and interesting material LaBaMn$_2$O$_6$ shows a transition from the paramagnetic metal to the FM phase at $T_C\simeq$ 330 K\@.\cite{ueda04,akahoshi03,miyauchi10,nakajima03,kawasaki09}
The transition temperature is moderately enhanced with respect to 280 K in La$_{0.5}$Ba$_{0.5}$MnO$_3$\@.
The neutron diffraction and NMR measurements revealed that part of the FM phase transforms to the AFI(CE) phase below $\sim$ 200 K and that the AFI(CE) phase coexists with the FM phase in the ground state.\cite{nakajima03,kawasaki06}
A similar coexistence was previously observed in $R_{1-x}A_x$MnO$_3$ and attributed to the A-site disorder or fluctuation of composition.
The observation of the phase separation in the A-site ordered LaBaMn$_2$O$_6$ reveals that this is not due to disorder at the A site, {\it i.e.}, phase separation and coexistence of ferro- and antiferromagnetic phases are intrinsic phenomena in the perovskite manganites.
Previous NMR results showed that the microscopic magnetic properties of the FM phase are very similar to those of the disordered La$_{0.5}$Ba$_{0.5}$MnO$_3$\@.\cite{kawasaki06}
While the transition temperature is moderately enhanced, the magnetic properties of the FM phase are not affected by the order or disorder at the A site in LaBaMn$_2$O$_6$\@.

Given the puzzling state of affairs regarding the AFI(CE) and the FM phases in the A-site ordered manganites, we have performed muon-spin-relaxation ($\mu$SR) measurements on YBaMn$_2$O$_6$ and LaBaMn$_2$O$_6$\@.
We stress that these two compounds provide a unique opportunity to study different aspects of the A-site order, which result in dramatic changes of the physical properties in the Y-compound, and a uniform ground state may be attributed directly to the A-site order. 
As explained above the situation is very different for the La-compound. 

Our results reveal an inhomogeneous magnetic FM phase for the A-site ordered LaBaMn$_2$O$_6$\@.
This is indicated by the absence of muon spin oscillation in the $\mu$SR time spectra and the significant deviation from a single exponential form in the relaxation.
The spatially inhomogeneous spin dynamics in the FM phase may be naturally associated with intimate intermixing of different magnetic phases, such as in the case of the Griffiths phase. 
Within this magnetically inhomogeneous phase, an antiferromagnetic state develops below 150 K, which displays well defined static internal magnetic fields, but reaches only 30\% of the volume fraction at low temperatures. 
In the AFI(CE) state of YBaMn$_2$O$_6$, the magnetic properties extracted from the $\mu$SR results agree with the expectations for a charge-ordered antiferromagnet, {\it i.e.}, the observed muon spin oscillating frequencies are consistent with the checker-board-type charge order and the antiferromagnetic spin order proposed by the neutron diffraction study, where $e_g$-electrons of Mn$^{3+}$ ions are well localized.

\section{Experimental procedure}

Powders of {\it R}BaMn$_2$O$_{5+\delta}$ ({\it R} = Y and La) were obtained by a solid-state reaction of {\it R}$_2$O$_3$, BaCO$_3$ and MnO$_2$\@.
After repeating the sintering process in pure Ar gas, the obtained ceramics were annealed in flowing O$_2$ gas, resulting in the A-site ordered perovskite {\it R}BaMn$_2$O$_6$. 
The details of the preparation method are described in the literature.\cite{ueda04}
The crystal structure and magnetic properties of both samples were measured and found to be consistent with the published data.\cite{ueda04}

The samples for the present $\mu$SR measurements were used in the previous $^{55}$Mn-NMR experiments.\cite{kawasaki06,kawasaki09}
In the $^{55}$Mn-NMR spectrum, one can clearly distinguish between the A-site ordered LaBaMn$_2$O$_6$ and the disordered La$_{0.5}$Ba$_{0.5}$MnO$_3$ from their resonance frequencies.\cite{kawasaki06}
We confirmed that the majority of the material is A-site ordered for both samples; the degree of A-site order is about 92\% for LaBaMn$_2$O$_6$\@.
As for YBaMn$_2$O$_6$, we did not observe $^{55}$Mn-NMR signal originating from impurity or secondary phases,\cite{kawasaki09} although the X-ray powder diffraction detected minor impurity of BaMnO$_{3-\delta}$ (less than 1\%).\cite{nakajima04jssc}

The $\mu$SR data for YBaMn$_2$O$_6$ have been taken by using surface muon at the M20 muon channel at TRIUMF (Vancouver, Canada) and on GPS at PSI (Villigen, Switzerland). For the $\mu$SR measurements on LaBaMn$_2$O$_6$, we have employed the M20 muon channel at TRIUMF and the $\pi$A muon channel at KEK (Tukuba, Japan).
All the $\mu$SR data presented here have been measured in zero field (ZF).

\begin{figure}[tb]
\centering
\includegraphics[width=7.5cm]{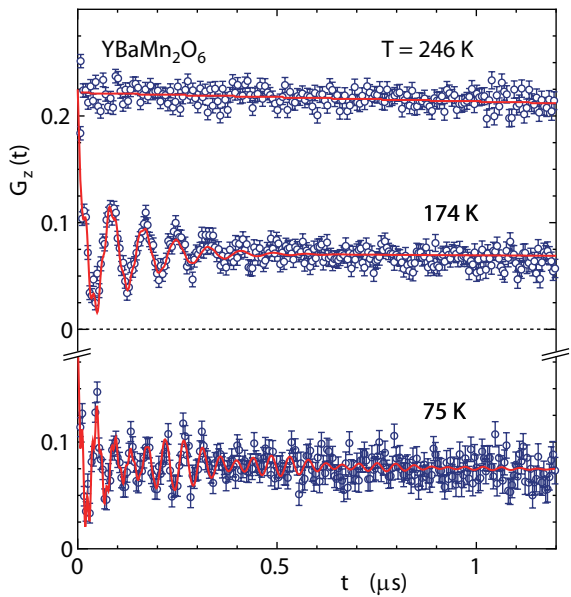}
\caption{(Color online) ZF-$\mu$SR time spectra of YBaMn$_2$O$_6$ at 246, 174 and 75 K\@.}
\end{figure}

\section{Results and discussion}
\subsection{YBaMn$_2$O$_6$}

We show three examples of the ZF-$\mu$SR time spectra of YBaMn$_2$O$_6$ above and below $T_N$ = 195 K in Fig.\ 2. 
A spontaneous Larmor precession is observed in the data below $T_N$, characteristic of a static magnetically ordered phase.
The Fourier transform FFT of the time spectra below $T_N$ are shown in Fig.\ 3, where four well-resolved frequencies are evident at low temperatures.
The time spectra above $T_N$ were fitted using $G_z(t)=A\exp(-\lambda t)$, where $A$ and $\lambda$ are the initial total asymmetry and the relaxation rate, respectively.
Below $T_N$, the functional form of the ZF-$\mu$SR time spectra includes oscillating terms
\begin{eqnarray}
G_z(t) & = & A_1\exp(-\lambda_1t) \nonumber \\
 & + & \sum_{i} A_{2,i}\exp(-\lambda_{2,i}t)\cos(2\pi\nu_i t+\phi),
\end{eqnarray}
describing the effect of the  static internal fields.
$A_{2,i}\ (\sum_i A_{2,i} = A-A_1)$ and $\nu_i$ are the $i$-th oscillating components of amplitude and frequency, respectively.
$\phi$ is the initial phase, the same for all the oscillating terms.
The associated relaxation rates, longitudinal and transversal, are $\lambda_1$ and $\lambda_{2,i}$, respectively.
The red solid lines in Fig.\ 2 represent the best fits to the data by using Eq.~(1)\@.

\begin{figure}[tb]
\centering
\includegraphics[width=7.0cm]{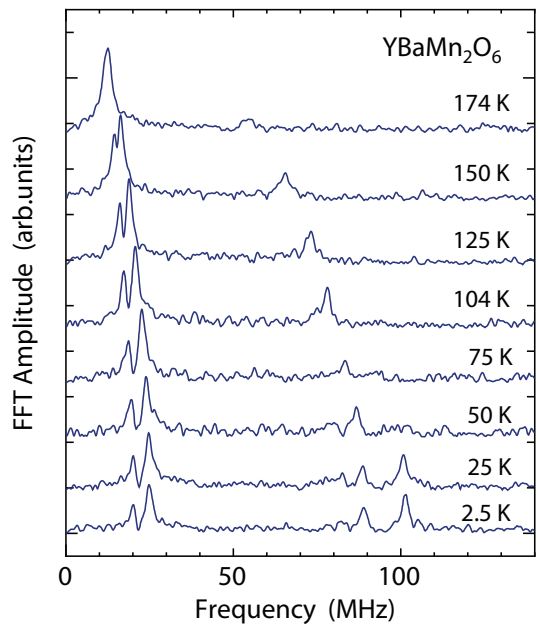}
\caption{(Color online) Fourier transform FFT of the time spectra below $T_N$ in YBaMn$_2$O$_6$\@.}
\end{figure}

The parameters $A$, $A_{\rm 1}$ (a), $\lambda$, $\lambda_1$ (b) and $\nu_i$ (c) obtained for various temperatures are summarized in Fig.\ 4, where $A$ and $\lambda$ ($A_1$ and $\lambda_1$) correspond to $T>T_N$ ($T<T_N$)\@.
A slight but clear hysteretic behavior in these parameters for the data measured on cooling (filled symbols) and on heating (open symbols) reveals first-order features of the antiferromagnetic phase transition.
The situation, however, is complicated because the transition also reveals critical fluctuations  (see Fig. 4b) which are normally associated with second-order type phase transitions. 
All this points a weak-first order type phase transition.\cite{lazuta06} 
We note that the first-order features may be related to a possible structural transition.
The antiferromagnetic order is accompanied by a change in the COO superstructure.\cite{kageyama03}

The initial total asymmetry $A \simeq 0.22$ has a very weak temperature dependence above $T_N$\@.
$A_1$, representing the signal from muons polarized along the direction of the internal field, decreases due to the antiferromagnetic spin order just below $T_N$\@.
We define the transition temperature as $T_N=$ 183 K and 173 K for warming and cooling process, respectively, from the inflection point of $A_1$. 
The ratio $A_1/A \simeq 1/3$  is expected if all the implanted muons are exposed to static local fields in a polycrystalline sample.
$A_1/A$ is, however, slightly larger than 1/3 below $T_N$ and reaches 1/3 only below 50 K\@.
This result suggests that a very small fraction of the sample, most probably an minority phase BaMnO$_{3-\delta}$ or Y$_{0.5}$Ba$_{0.5}$MnO$_3$, orders magnetically below $\sim$ 50 K\@.
Therefore, the muon spin oscillation with $\nu_4$ = 101.1 MHz (see Fig.\ 3 and Fig.\ 4(c)), that appears only below 50 K, is considered to be extrinsic.
The existence of a minority phase is also indicated by an anomaly in $\chi(T)$ at the same temperature.\cite{nakajima04}
The broad peak in the relaxation rate just below $T_N$ may be due to a critical slowing down of Mn magnetic moments accompanying the antiferromagnetic order.

Below $T_N$ one detects in the $\mu$SR time-spectra oscillations with frequencies $\nu_1$ = 20.2 MHz, $\nu_2$ = 24.8 MHz and $\nu_3$ = 88.9 MHz\@.
They reveal quasi-static internal magnetic fields of $B = (2\pi/\gamma)\nu$, where $\gamma = 2\pi\times135.5$ MHz/T is the muon gyromagnetic ratio and $B$ is a local static magnetic field at muon site.
The inferred local fields are 0.15 T, 0.18 T and 0.66 T at 2.5 K, respectively.
The high values of the frequencies just below $T_N$, $\nu_i(T_N)/\nu_i(0)\simeq0.6$, reveal a first-order type features of the magnetic transition.

\begin{figure}[tb]
\centering
\includegraphics[width=7.5cm]{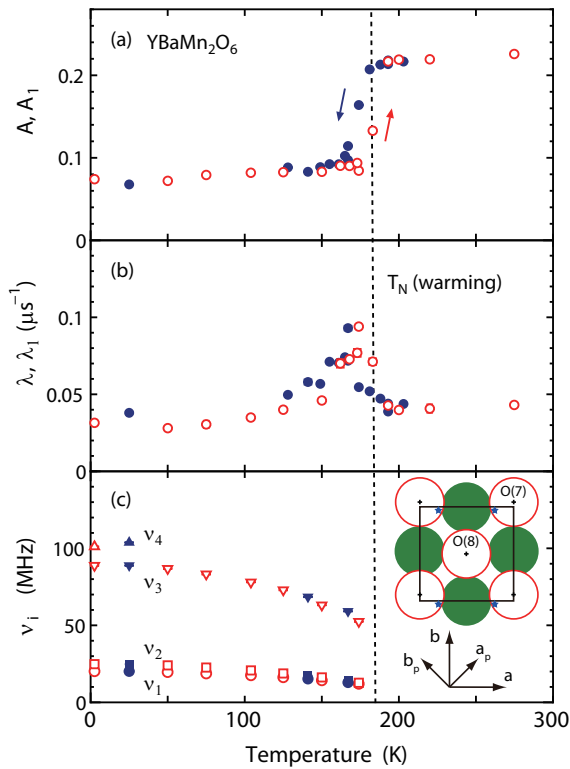}
\caption{(Color online) Temperature dependence of $A$ and $A_{\rm 1}$ (a), $\lambda$ and $\lambda_1$ (b) and $\nu_i$ (c) in YBaMn$_2$O$_6$\@. Filled and open symbols indicate the data measured on cooling and warming, respectively. Inset shows the BaO plane ($z=0.5$), where green filled circles, red open circles and blue stars represent Ba, O and muon sites, respectively.}
\end{figure}

The observation of three precession frequencies indicates multiple  magnetically inequivalent stopping sites for the muon.
Considering that dipolar fields dominate the local field at the muon site in LaMnO$_3$,\cite{heffner01,guidi01} we estimated the fields for possible muon sites about 1\AA\  away from oxygen ions and in the large empty space in the MnO$_2$ plane, and adopting the crystalline and the magnetic structure with the CE-type ordering in the MnO$_2$ plane with the $\alpha\alpha\beta\beta$-type stacking along the $c$ axis.\cite{nakajima04jssc,kageyama03}
We propose a likely muon site at $(0.8, 0.96, 0.5)$, about 1.2\AA\  away from the oxygen positions in the BaO plane, where the estimated dipolar fields agree roughly with the observed local fields.
The position of the muon site is indicated by blue stars in the inset of Fig.~4(c), where all the plotted stars are crystallographically equivalent.
It is close to the muon site for LaMnO$_3$,\cite{guidi01,heffner01} but not exactly the same.
One may not expect to have exactly the same muon sites, because the O-ion displacements from their ideal positions are much smaller in YBaMn$_2$O$_6$ than those in LaMnO$_3$\@.
In addition, the spatial distribution of electrostatic potential are expected to be different each other, because Mn$^{3+}$ and Mn$^{4+}$ have the checker board type ordering in the MnO$_2$ plane of YBaMn$_2$O$_6$, while the valence of Mn ion is uniform for LaMnO$_3$\@.
We stress that not only the magnitude of the estimated fields of 0.208, 0.232, 0.235, and 0.580 T at the muon sites agree with the experimental results but also the relative intensities are consistent with the observed intensity in the Fourier spectra (See Fig.\ 3).
In particular, the intensity at $\nu_2$ is twice as large as those at $\nu_1$ and $\nu_3$.
A detailed discussion on the muon site derivation is given in the appendix.

In the case of LaMnO$_3$, the transferred hyperfine field vanishes by symmetry considerations of the A-type magnetic structure (ferromagnetic planes are stacked antiferromagnetically along $c$-axis) with a tiny canting.\cite{guidi01}
For YBaMn$_2$O$_6$, symmetry arguments do not imply a vanishing transferred hyperfine field in the BaO plane.
Nevertheless, the estimated field at the muon site $(0.8, 0.96, 0.5)$ is consistent with the experimental results and the site is close to that in LaMnO$_3$\@.
We conclude, therefore, that the dipolar contribution to the local field is dominant in YBaMn$_2$O$_6$\@.
The transferred hyperfine field may account for the small difference between the estimated dipolar fields and experimentally observed values.

\subsection{LaBaMn$_2$O$_6$}

Figure 5 shows the $\mu$SR time spectra of LaBaMn$_2$O$_6$ in zero field at 324, 300, 240 and 160 K\@.
In the FM phase, below $T_C$, no clear muon spin oscillation is observed.
The onset of the magnetic order is indicated by a decrease in the initial asymmetry and the deviation from a single exponential form in the relaxation curve.
The time-spectrum data have been fitted using a stretched exponential model
\begin{equation}
G_z(t) = A\exp(-\lambda t)^n,
\end{equation}
where $A$ is an initial asymmetry and $\lambda$ is a longitudinal muon spin relaxation rate and $n$ is a stretched exponent.
A stretched exponent $n\simeq1$ reveals a small distribution of the relaxation rates, while smaller values, $n<1$, suggests a distribution of the rates.
The best fits are represented by the red solid lines in Fig.~5\@.

The temperature dependence of $A$, and $\lambda$ and $n$ are summarized in Figs.\ 6 (a), (b) and (c), respectively.
Here, the dotted line represents the Curie temperature, $T_C \simeq 300$ K,  defined as the inflection point of $A(T)$\@.
Above $T_c$, $A\simeq0.22$ has a weak temperature dependence.
With decreasing temperature, $A$ starts to decrease at 330 K and reaches about one third of the value in the paramagnetic metal phase below 250 K\@.
The unobserved initial asymmetry at low temperatures in Figs.\ 5 and 6 ($2/3$ of the total) is interpreted as evidence for a very fast relaxing component (probably transversal).
We only observe the relatively slow relaxing component (probably longitudinal).
Another striking feature of our data is the large inhomogeneous broadening of the magnetic transition which is discussed  below.
One third of the initial asymmetry in the magnetically ordered state represents the signal from muons polarized along the direction of the internal field in a powder sample. 
Since in our case this value corresponds to one third of the total asymmetry in the paramagnetic state, our data show that the volume fraction of magnetic phase is almost 100\%.

\begin{figure}[tb]
\centering
\includegraphics[width=7.5cm]{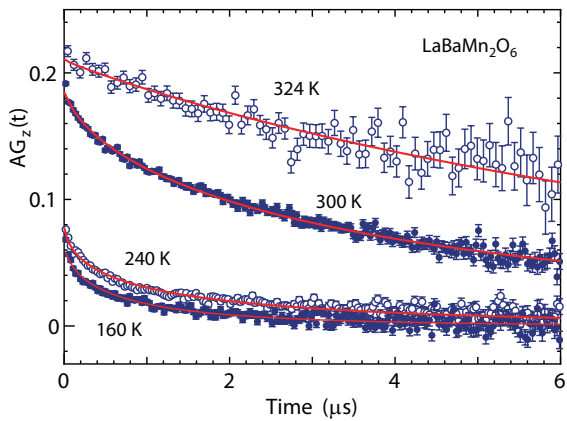}
\caption{(Color online) Time spectra of LaBaMn$_2$O$_6$ at 324, 300, 240 and 160 K in zero magnetic field. The solid curves indicate the best fit by the stretched exponential model for respective temperatures.}
\end{figure}

The temperature dependence of $\lambda$ reveals a peak at 280 K, somewhat below $T_C$, and a broad maximum at  125 K\@.
The peak at 280 K may be associated to a critical slowing down of Mn magnetic moments usually accompanying the onset of ferromagnetic order, although the peak temperature is slightly lower than $T_C$\@.
We note, however, that this scenario has a difficulty. 
Namely, here the transition region is very broad and in such case it is not clear that one can detect critical slowing down. 
Therefore, one should consider a different scenario. 
Namely, between 250 and 300 K, part of the fast relaxing (at low temperatures) component may be in fact observed in the $\mu$SR time signal.
The inferred  relaxation rate then may be the result of an interplay between the increase in the fast (transversal) relaxation rate and the gradual decrease in its weight.
The very broad signal centered around 125 K, ranging from 50 K to 200 K, may be due to the change in a magnetic structure associated with the appearance of AFI(CE) phase in the FM phase below $\sim$ 200 K\@.

At and below $T_C$, the relaxation curve changes from an exponential ($n = 1$) form in the paramagnetic metal phase to a stretched exponential ($n = 1/2$) in the FM phase, suggesting a large distribution of the relaxation rates.
With decreasing temperature below 150 K, $n$ decreases further and reaches about 0.2 at 2.5 K\@.

\begin{figure}[tb]
\centering
\includegraphics[width=7.5cm]{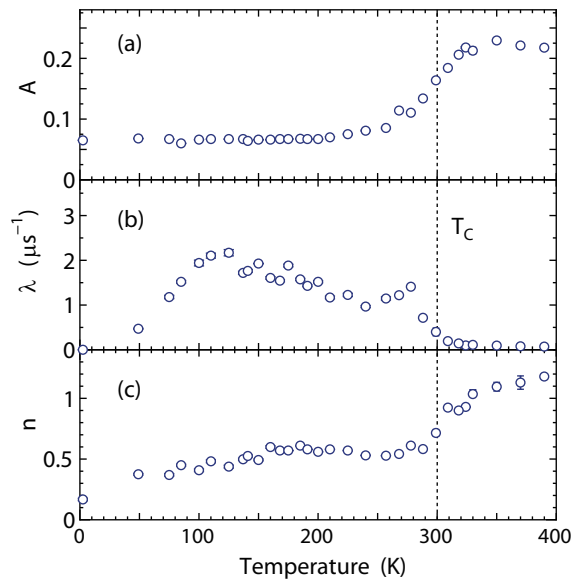}
\caption{(Color online) Temperature dependence of $A$ (a), $\lambda$ (b) and $n$ (c) for LaBaMn$_2$O$_6$ data measured at zero magnetic field.}
\end{figure}

The absence of muon spin oscillations, as well as the significant deviation from the single exponential form in the time spectra, suggests a distribution of the relaxation rates in the FM phase of LaBaMn$_2$O$_6$\@.
This distribution may originate from spatially inhomogeneous spin dynamics, as observed by $\mu$SR experiments in the A-site disordered ferromagnetic manganites, such as  La$_{1-x}$Ca$_{x}$MnO$_3$ ($x$ = 0.18, 0.33 and 0.375) and La$_{0.85}$Sr$_{0.15}$MnO$_3$.\cite{heffner01,heffner96,barsov00,heffner00,heffner00a, heffner03}
These facts are consistent with the two-phase interpretation of a percolation theory.\cite{moreo99}
This result suggests that the microscopic magnetic properties of the FM phase in LaBaMn$_2$O$_6$ are similar to those in the A-site disordered ferromagnetic manganites, irrespectively of the existence of order/disorder at the A site.
This finding adds support to the conclusion obtained by $^{55}$Mn-NMR measurements that the magnetic properties of the FM phase in LaBaMn$_2$O$_6$ are similar to those of the disordered form La$_{0.5}$Ba$_{0.5}$MnO$_3$\@.\cite{kawasaki06}

Spatially inhomogeneous spin dynamics in the FM phase occurs in a variety of very different systems and go under the denomination of ``Griffiths phase''.\cite{lazuta06}
Not always it is possible to identify the mechanism that triggers the inhomogeneous phase. 
In most cases the only common characteristic of these diverse systems is the presence of strong electronic correlations, although in some particular cases chemical inhomogeneities have been named as a possible cause. 
For our materials the degree of A-site order is estimated to be about 92\% for the measured sample.\cite{kawasaki06}
The mismatch of lattice parameters $a$, $b$ between the A-site ordered and the disordered structures is very small, as it is the case for the Ruddlesden-Popper series of layered perovskites.
The coexistence of the two, therefore, may take the form of intergrowth, where the layers of the minority structure are intercalated in the other, as has been identified in the double-layer manganites (La$_{1-z}$Pr$_z$)$_{1.2}$Sr$_{1.8}$Mn$_2$O$_7$\@.\cite{allodi08}
This type of random arrangements of the layers may be the origin of inhomogeneous spin dynamics in the FM phase of LaBaMn$_2$O$_6$ and possible Griffiths phase phenomenon.\cite{pramanik10}
We cannot make a definite statement in our case.

\begin{figure}[tb]
\centering
\includegraphics[width=7.5cm]{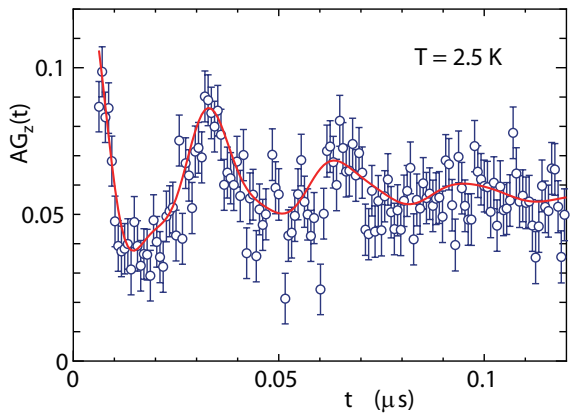}
\caption{(Color online) Time spectrum for LaBaMn$_2$O$_6$ at 2.5 K in zero external field. The red solid curve is the best fit of Eq.\ (1), indicating two components of muon oscillation frequency, $\nu_1$ = 69.6 MHz and $\nu_2$ = 31.6 MHz\@.}
\end{figure}

There is a modest, but clear decrease in $n$ below 150 K\@.
In fact this modest change implies a substantial redistribution of the relaxation rates, and is probably due to the appearance of the AFI(CE) phase inside the FM one.
The spin dynamics in the AFI(CE) phase may differ from that in the FM phase, resulting in a large distribution of $\lambda$.
The appearance of the AFI(CE) phase is indicated by the muon spin precession below around 125 K\@.
Figure 7 shows the time spectrum at 2.5 K with the best fit by using Eq.\ (1), indicating two components of muon precession frequency of  $\nu_1$ = 69.6 MHz and $\nu_2$ = 31.6 MHz\@.
The first term of Eq.\ (1) represents the signal from muons polarized along the direction of the internal field in the FM and AFI(CE) phases (longitudinal signal) and the oscillatory terms represent the transversal signal in the AFI(CE) phase. 

The temperature dependence of $\nu_1$ and $\nu_2$ is shown in Fig.\ 8\@.
$\nu_1$ and $\nu_2$, being approximately temperature independent and still high at 125 K\@.
This indicates that the antiferromagnetic spin structure arises from the ferromagnetic phase through a first-order type transition (spin re-orientation).
This result is consistent with the temperature independent $^{55}$Mn-NMR frequency in the AFI(CE) phase.\cite{kawasaki06}

One may roughly estimate the volume fraction of the AFI(CE) phase at 2.5 K as follows. 
The sum of oscillating amplitudes, $A_{2,1}+A_{2,2}= 0.041$, may be considered to be 2/3 of the initial asymmetry from muons in the AFI(CE) phase. 
Thus, the volume fraction of the AFI(CE) phase is estimated to be $(3/2)(A_{2,1}+A_{2,2})/A \sim 30$\%, where $A\sim0.22$ is the initial asymmetry at high temperature above $T_C$\@.
This estimated value is in good agreement with the value of 30\% obtained from the magnetization curve.\cite{nakajima03}

\begin{figure}[t]
\centering
\includegraphics[width=7.5cm]{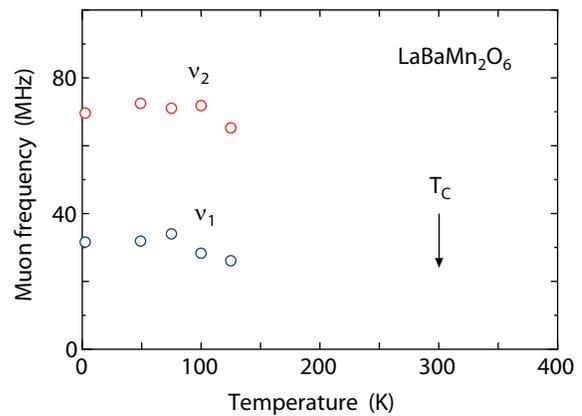}
\caption{Temperature dependence of the zero-field muon frequencies for LaBaMn$_2$O$_6$\@.}
\end{figure}

From considerations of possible muon sites, we concluded that most likely the muons stop at the position $(0.755, 1/2, 1/2)$ for LaBaMn$_2$O$_6$, see appendix for details. 
As explained in the appendix, for this site we estimated the dipolar field assuming a uniform ferromagnetic state, and obtained 0.38 T corresponding to 51.5 MHz\@.
 This is clearly within our range of detection.
But the fact that no muon spin oscillations are observed, adds support to the claim that the inhomogeneity of the FM phase is in fact intrinsic (as opposed to the scenario where well defined $\mu$SR frequencies are simply too high to be observed).

\begin{figure}[t]
\centering
\includegraphics[width=7.5cm]{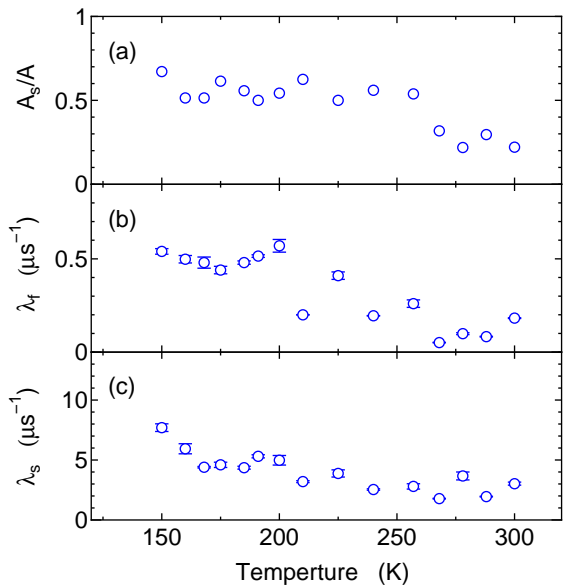}
\caption{Temperature dependence of fractional amplitude of $A_s$, {\it i.e.} $A_s/A$ (a), $\lambda_f$ (b), and $\lambda_s$ (c) in the two-exponential model for the FM phase of LaBaMn$_2$O$_6$\@.}
\end{figure}

Finally, we present an alternative approach for the analysis of the data between 150 K $< T < T_C$\@. 
The two-exponential model takes into account a two-phase percolation proposed for ferromagnetic manganites below $T_C$.\cite{moreo99} 
It leaves out some of the subtleties, but it is simple to understand the meaning of the fitting parameters. 
Here, the time-spectrum data is expressed by
\[
G_z(t)=A_f\exp(-\lambda_f t)+A_s\exp(-\lambda_s t),
\]
where $\lambda_f$ and $\lambda_s$ are the relaxation rates for two exponentials, respectively. 
$A_f$ and $A_s$ are the corresponding $\mu$SR amplitudes.
Here, $A=A_f+A_s$ and $\lambda_f < \lambda_s$.
The smaller value of the relaxation rate $\lambda_f$ indicates the faster fluctuation rate of local field at the muon site.

In Fig~9, we show the temperature dependence of fractional amplitude of $A_s$, {\it i.e.} $A_s/A$ (a), $\lambda_f$ (b), and $\lambda_s$ (c) in the two-exponential model for the FM phase of LaBaMn$_2$O$_6$\@.
The $\lambda_s$ component appears around below $T_C\simeq300$ K and its amplitude $A_s$ increases ($A_f$ decreases) upon cooling.
Above $T_C$, $n\simeq1$ in the stretched exponential model indicates that the relaxation curve has a single exponential form, {\it i.e.}, $A_s\simeq0$ (see Fig. 6(c)). 
$A_s$ remains approximately independent of temperature below around 250 K, where the observed asymmetry is close to $1/3$ of the total.
Both values of the relaxation rates $\lambda_f$ and $\lambda_s$ become larger upon cooling down to 150 K\@.

The relaxation rate $\lambda_s$ is one order of magnitude larger than $\lambda_f$.
Similar results in the two-exponential model have been reported for the ferromagnetic conductor La$_{0.67}$Ca$_{0.33}$MnO$_3$ and the insulating ferromagnet La$_{0.82}$Ca$_{0.18}$MnO$_3$.\cite{heffner00,heffner01}
Heffner {\it et al.} associated the relaxations of $\lambda_f$ component  with overdamped spin waves, characteristic of a disordered ferromagnetic metal, and that of $\lambda_s$ component with relatively insulating regions of the sample.
The damping of spin waves may be caused by an inherent disorder due to the random replacement of La atoms by Ca.
In our case, there is no such a random replacement, but similar mechanism may be also responsible for the observed relaxation of $\lambda_f$ component due to possible random arrangements of the layers as discussed above.

In the ferromagnetic conductor La$_{0.67}$Ca$_{0.33}$MnO$_3$ $A_s$ decreases ($A_f$ increases) upon cooling below $T_C$, while $A_s$ is almost independent of temperature below around 170 K for the insulating ferromagnet La$_{0.82}$Ca$_{0.18}$MnO$_3$ that shows a transition to a CO state below around 60 K.\cite{heffner00,heffner01}
The temperature independent $A_s$ below around 250 K for LaBaMn$_2$O$_6$, which is qualitatively similar to that for La$_{0.82}$Ca$_{0.18}$MnO$_3$, indicates that the relatively insulating region survives even in the FM phase.
The weak metallic nature of the FM phase in LaBaMn$_2$O$_6$ may be consistent with the appearance of AFI(CE) phase inside the FM phase at low temperature.



\section{conclusion}
In conclusion, our zero-field $\mu$SR data reveal striking differences in the magnetically ordered  phases of YBaMn$_2$O$_6$ and LaBaMn$_2$O$_6$\@.
For the former, a homogeneous state is observed, as expected for an A-site ordered manganite.
For the latter, unexpectedly one finds inhomogeneous magnetic phases below room temperature.

It is not clear what triggers these important differences in our two materials.
On the other hand, the A-site order triggers important changes in the thermal and transport properties of the Y compound, but not in the La-compound.
The inhomogeneous magnetic phase of the La compound is reminiscent of the findings in (La$_{1-z}$Pr$_z$)$_{1.2}$Sr$_{1.8}$Mn$_2$O$_7$, where an inhomogeneous magnetic phase is thought to be triggered by peculiar stacking faults leading to inter-growth of different magnetic layers. 
But this mechanism does not seem to apply for the Y-compound.


\section*{appendix}
We consider a possible muon site in YBaMn$_2$O$_6$ on the basis of previous works in LaMnO$_3$\@.
The crystal structure of LaMnO$_3$ has an orthorhombic cell with space group $Pnmb$, where the Mn ions are located at the (0, 0.5, 0) positions.\cite{huang97}
Cestelli Guidi {\it et al.}\ have proposed two muon sites; (0.389, 0.937, 0.25) and (0.5, 0.5 ,0) for LaMnO$_3$\@.\cite{guidi01}
The first site, close to the Holzschuh site found for orthoferrites,\cite{holzschuh} is about 1.1\AA\ distant from an apical oxygen and farthest away from positive La ions in the LaO plane.
The second site is located in the large empty space at the center of the MnO$_2$ plane.
Independently, Heffner {\it et al.} have reported similar muon sites.\cite{heffner01}

The crystal structure of YBaMn$_2$O$_6$ has a monoclinic cell with space group $P3$\@.
There are two Mn sites, trivalent Mn(1) at  (0.018, 0.005, 0.242) and tetravalent Mn(2) at (0.481, 0.518, 0.244), and eight oxygen sites, O(1) $\sim$ O(8) in the unit cell at 350 K\@.\cite{nakajima04jssc}
Muons bound to apical oxygens, O(1) and O(2) in the YO plane and O(7) and O(8) in the BaO plane of YBaMn$_2$O$_6$, may be closely related to the first muon site in LaMnO$_3$, while muons bound to planar oxygens O(3) $\sim$ O(6) in the MnO$_2$ plane may be closely related to the second muon site.

Considering that dipolar fields dominate the local field at the muon site in LaMnO$_3$,\cite{heffner01,guidi01} we have estimated the fields at about 1\AA\  apart from the eight oxygens and in the large empty space around (0, 0.5, 0.25) in the MnO$_2$ plane of YBaMn$_2$O$_6$\@.
For this purpose we adopted (i) the structural parameters at 350 K proposed by X-ray and neutron powder diffractions,\cite{nakajima04jssc} (ii) the CE-type magnetic structure in the MnO$_2$ plane with the $\alpha\alpha\beta\beta$-type stacking along the $c$ axis\cite{kageyama03} and (iii) 2.65$\mu_{\rm B}$ for Mn$^{4+}$ ion and 3.89$\mu_{\rm B}$ for Mn$^{3+}$ ion as reported for CaMnO$_3$ and for LaMnO$_3$ from neutron diffraction, respectively.\cite{wollan55}
Although this assumption changes slightly the estimated values of dipolar fields with respect to a preliminary report,\cite{kawasaki09} it does not affect the conclusion respect to the muon site.

We found a possible muon site at $(0.8, 0.96, 0.5)$, being in rough agreement with the observed local fields of 0.15, 0.18 and 0.66 T\@.
This position is about 1.2\AA\  away from O(7) in the BaO plane and indicated by blue stars in the inset of Fig.~4(c), where all the plotted stars are crystallographically equivalent.
It is close to the first muon site proposed in LaMnO$_3$, but not exactly the same.
One may not expect to have exactly the same muon sites, because the O-ion displacements from their ideal positions are much smaller in YBaMn$_2$O$_6$ than those in LaMnO$_3$\@. 
In addition, the spatial distribution of electrostatic potential are expected to be different each other, because Mn$^{3+}$ and Mn$^{4+}$ have the checker board type ordering in the MnO$_2$ plane of YBaMn$_2$O$_6$, while the valence of Mn ion is uniform for LaMnO$_3$\@.
For this muon site, we find four magnetically inequivalent muon sites with the dipolar fields of 0.208, 0.232, 0.235, and 0.580 T along the $b$-axis.
The multiple magnetically inequivalent sites arise from the CE-type magnetic structure in the MnO$_2$ plane with the large magnetic unit cell.
The transferred hyperfine field may account for the small difference between the estimated dipolar fields and the experiments as mentioned in the section III.A\@.
Our estimates are also consistent with the observed intensity in the Fourier transformed spectrum where intensity at $\nu_2$ is twice as large as those at $\nu_1$ and $\nu_3$.

Regarding our chosen muon site, it seems very likely that positive muons stop in the BaO plane by two reasons; (i) Ba$^{2+}$ has smaller positive charge than Y$^{3+}$, (ii) the average distance between BaO and MnO$_2$ planes is larger by 25\% than that between YO and MnO$_2$ planes due to the distortion of MnO$_6$ octahedra.\cite{nakajima04jssc}
In addition, muons may favor a position near O(7) rather than a position near O(8) in the BaO plane, because the nearest Mn ion to O(7) is trivalent Mn(1) while that to O(8) is tetravalent Mn(2)\@.

Regarding LaBaMn$_2$O$_6$, it seems reasonable to assume its muon site to be the position related to Holzschuh site or its symmetric replica for orthoferrites.\cite{holzschuh}
In the tetragonal cell of LaBaMn$_2$O$_6$, there are three oxygen sites; O(1) at (1/2, 1/2, 0) in the LaO plane, O(2) at (1/2, 0, 0.2373) in the MnO$_2$ plane and O(3) at (1/2, 1/2, 1/2) in the BaO plane at 400 K.\cite{nakajima03}
There are no O-ion deviations from the ideal positions in BaO plane of LaBaMn$_2$O$_6$\@.
We propose that the muon lies at (0.755, 1/2, 1/2) in the BaO plane, 1 \AA \ away from O(3) towards the midpoint between two adjacent Ba ions.
We estimated the dipolar sums over a large spherical domain centered at the proposed muon site in the FM phase, assuming a uniform ferromagnetic state.
In the estimation, we used the structural parameters at 400 K and the Mn magnetic moment 3.0 $\mu_B$ from the magnetization in 5 T, where the AFI(CE) phase is suppressed by the magnetic field.\cite{nakajima03}
The estimated dipolar field is 0.38 T, corresponding to the muon frequency of 51.5 MHz\@.

\ \ \ \ \ \ \ \ \ \ \ \ \ \ \ \ \ \ \ \ \ \ \ \ \ \ \ \ \ \ \ \ \ \ \ \ \ \ \ \ \ \ \ \ \ \ \ \ \ \ \ \ \ \ \ \ \ \ \ \ \ \ \ \ \ \ \ \ \ \ \ \ \ \ \ \ \ \ \ \ \ \ \ \ \ \ \ \ \ \ \ \ \ \ \ \ \ \ \ \ \ \ \ \ \ \ \ \

\end{document}